\documentclass[twocolumn,astrosymb,times]{aastex701}
\usepackage{rotating}
\usepackage{booktabs}
\usepackage{textcomp,gensymb}
\usepackage{tabularx}
\usepackage{amsmath}
\usepackage{textgreek}

\shortauthors{P\'erez Paolino et al.}

\begin{document}

\title{Testing Pre-main-sequence Evolutionary Models with Gaia and SPHEREx Observations of Canis Major North}

\correspondingauthor{Facundo P\'erez Paolino}

\author[0000-0002-4128-7867, gname=Facundo, sname=P\'erez Paolino]{Facundo P\'erez Paolino}
\affiliation{Department of Astronomy, California Institute of Technology, 1216 East California Blvd, Pasadena, CA 91125, USA}
\email[show]{fperezpa@caltech.edu}

\author[0000-0002-6549-9792]{Ronan Kerr}
\affiliation{Dunlap Institute for Astronomy \& Astrophysics, University of Toronto,
Toronto, ON M5S 3H4, Canada\\}
\email{ronan.kerr@utoronto.ca}

\author[0000-0002-2592-9612, gname=Jonathan, sname=Gagn\'e]{Jonathan Gagn\'e}
\affiliation{Plan\'etarium de Montr\'eal, Espace pour la Vie, 4801 av. Pierre-de Coubertin, Montr\'eal, Qu\'ebec H1V 3V4, Canada}
\affiliation{Trottier Institute for Research on Exoplanets, Universit\'e de Montr\'eal, D\'epartement de Physique, C.P. 6128 Succ. Centre-ville, Montr\'eal, Qu\'ebec H3C 3J7, Canada}
\email{jonathan.gagne@montreal.ca}


\begin{abstract}
Pre-main-sequence (PMS) evolutionary models can assign different ages across stellar mass within the same population. We use joint Gaia and SPHEREx spectral energy distributions to infer effective temperatures, radii, and luminosities for members of Canis Major North (CMaN), a nearby association with independent lithium and kinematic age constraints near 23 Myr. The SPHEREx fluxes agree with 2MASS and WISE photometry to 3--6\%, and we anchor the absolute flux scale to Gaia, 2MASS, and WISE. Using the revised temperatures, we obtain a lithium age of $22.4^{+2.8}_{-1.5}$ Myr. For a 29-star PMS sample, population-wide HR-diagram fits yield $20.5^{+1.5}_{-1.4}$ Myr for unspotted SPOTS models, consistent with the external constraints, $18.4^{+1.4}_{-1.3}$ and $18.7^{+1.3}_{-1.2}$ Myr for the Feiden nonmagnetic and BHAC15 models, and substantially older ages of $32.2^{+2.3}_{-2.1}$ and $55^{+5}_{-5}$ Myr for the Feiden magnetic and PARSEC models. No grid returns a temperature-independent age. The Feiden, SPOTS, and BHAC15 models assign older ages to stars above $4000$ K, whereas PARSEC assigns increasingly older ages toward lower temperatures. The intrinsic luminosity dispersion is also mass dependent, approximately $0.14$ dex below $3450$ K but $0.05$ dex at intermediate temperatures. At a fixed age of 23.0 Myr, a single SPOTS filling factor cannot reproduce the width of the low-temperature sequence, whereas the empirical filling-factor distribution of young stars overpredicts it. Spotted isochrones can enclose every cool member only at ages of 37--50 Myr, inconsistent with the lithium and kinematic ages.
\end{abstract}

\section{Introduction}
Ages are the clock against which the history of star formation is reconstructed. Young associations retain the motions of their parent clouds, so their positions, velocities, and ages allow individual star-forming events to be traced back in space and time and related to one another \citep{dezeeuw99,Gagne21}. Testing scenarios such as triggered star formation \citep{Elmegreen77} requires ages accurate to a few Myr, an accuracy that remains difficult to reach for pre-main-sequence (PMS) stars.

Isochronal ages are by far the most widely available, owing to precise astrometry and photometry. Gaia parallaxes now allow placing the PMS populations of most associations within 1 kpc on the Hertzsprung--Russell diagram (HRD), and their relative ordering is robust: younger populations lie farther above the main sequence \citep{Kerr24}. This ordering has revealed spatial age gradients and possible sequences of star formation \citep{Pang21,Ratzenboeck2023,Posch25}. The absolute scale is far less certain, however. Different evolutionary models suggest ages that disagree by a factor of two or more \citep{Herczeg15}, and a single model often assigns different ages to stars of different mass in the same population. In Upper Sco, isochronal ages are typically 4--5 Myr for low-mass stars but 11--13 Myr for intermediate-mass stars \citep{Pecaut2016,Fang2017}. Part of this gradient can arise from star-formation history \citep{Fang2017} or from unresolved binaries \citep{SullivanKraus2021}, but part reflects the models themselves. Empirical sequences of benchmark associations provide one route to a self-consistent scale \citep{Bell12_pre-main-sequence_isochrones,Bell13_pre-main-sequence_isochrones,Herczeg15}.

Anchoring that scale requires age estimates that are independent of the HRD. Lithium depletion boundary (LDB) ages are largely insensitive to the input physics and the initial lithium abundance \citep{Burke04}, but they require spectroscopy of faint M stars and are therefore limited to nearby or well-populated groups \citep{Wood23}. Lithium depletion among warmer stars, on the other hand, yields an empirical age with larger intrinsic scatter \citep{Jeffries23}. Where LDB ages exist, they have been found to exceed the color--magnitude ages from the same models by a factor of two, a discrepancy reduced by models with magnetic inhibition of convection or starspots \citep{Binks2021}. Kinematic ages, obtained by tracing an expanding population back to its most compact configuration, avoid stellar physics altogether \citep{Luhman2023,MiretRoig24}. They measure, however, the time since the stars became unbound. Some associations show no coherent expansion \citep{WrightMamajek2018}, and traceback ages tend to fall several Myr below isochronal ages \citep{MiretRoig24}, plausibly because stars remain bound to their parent gas for some time after they form \citep{Kerr22a}.

PMS models differ in their treatment of convection, magnetic inhibition, starspots, atmospheric boundary conditions, and initial conditions. Each choice changes both the absolute age scale and how the predicted isochrone performs with temperature \citep{Soderblom2014,Somers2015,Flores2022}. Starspots, for example, block emergent flux and inflate radii, so unspotted models underestimate the masses and ages of spotted stars \citep{Somers2015,Cao2022, PerezPaolino2024}. Dynamical masses from eclipsing binaries and circumstellar disks show that no single prescription is correct at all masses: magnetic models reproduce dynamical masses above $\approx0.6\,M_\odot$ but not below \citep{Braun2021}, and in Upper Sco the model family and spot coverage that best match dynamical masses change with stellar mass \citep{Towner2025,Zallio2026}. A coeval population with independent lithium and kinematic ages tests both the absolute calibration of a grid and its mass dependence.

Canis Major North (CMaN; SCYA-65) provides such a test. Identified as a young Gaia population in the SPYGLASS survey \citep{Kerr2023}, it contains $\approx190$ candidate members with a combined mass of $\approx70\,M_\odot$, at a mean distance of $\approx180$ pc toward $(l,b)\approx(226^\circ,-8^\circ)$, and it shows no detectable kinematic or spatial substructure \citep{Kerr2025a}. \citet{Kerr2025a} measured a dynamical age of $23.7\pm4.9$ Myr and lithium-depletion ages of $26.3\pm2.0$ and $24.5\pm1.6$ Myr from two implementations of EAGLES. Their PARSEC and DSEP-Magnetic color--magnitude ages, $26.0\pm1.0$ and $23.0\pm1.1$ Myr, agree with these, whereas BHAC15 gave $13\pm0.3$ Myr. CMaN and the neighboring association Theia~72 trace back to a common formation site $\approx26$ Myr ago, and their asymmetric velocities suggest acceleration by an expanding bubble or a cloud--cloud collision \citep{Kerr26_spyglass_vii-b}. The age of CMaN therefore also dates the formation of that complex.

In this paper, we use joint Gaia and SPHEREx spectral energy distributions (SEDs), anchored to 2MASS and WISE photometry, to place CMaN members on the HRD and compare with five PMS evolutionary models. The Feiden magnetic/nonmagnetic and SPOTS spotted/unspotted calculations isolate the effects of magnetic inhibition and starspots, while PARSEC and BHAC15 \citep{Baraffe2015} provide widely used references. We revise the lithium age with the new temperatures, fit the HRD globally and in separate temperature intervals, and test whether heterogeneous spot coverage can reproduce the observed sequence at the lithium age. Sections~\ref{sec:obs}--\ref{sec:lithium_age} describe the observations, sample, SED fits, and lithium age; Sections~\ref{sec:hrd_models} and \ref{sec:fspot_forward_model} present the evolutionary-model and starspot tests; and Sections~\ref{sec:discussion} and \ref{sec:conclusions} discuss the results and summarize our conclusions.

\section{SPHEREx Observations and Sample Selection}
\label{sec:obs}

The Spectro-Photometer for the History of the Universe, Epoch of Reionization, and Ices Explorer (SPHEREx) is a NASA Explorer mission carrying out an all-sky near-infrared spectral survey \citep{Crill2020,Bock2026}. SPHEREx provides spectra spanning $0.75$ to $5.0~\mu{\rm m}$ in 102 spectral bands, with a resolving power ranging from $R\approx35$ to 130 and a native pixel scale of $6\farcs15$ \citep{Bock2026}. Spectrophotometry is obtained using linear variable filters mounted above six HgCdTe H2RG detector arrays. Because the central wavelength transmitted by each filter varies continuously with position across the focal plane, a source is measured at different wavelengths as repeated spacecraft pointings place it at successive detector positions \citep{Crill2020,Hui2026}. The resulting measurements form densely sampled, low-resolution spectral energy distributions rather than conventional slit or dispersive spectra.

Our targets were drawn from the CMaN membership catalog of \citet{Kerr2025a}. The SPHEREx extraction was run for the 146 candidates with spatial membership probability $P_{\rm spat}>0.5$, a deliberately inclusive list; the published spatial--photometric selection $P_{\rm sp}>0.5$ contains 118 of them. Membership of the stars used for the age analysis is assessed with the published probabilities in Section~\ref{sec:cuts}. We retained targets with available SPHEREx observations and Gaia DR3 astrometry and photometry for SED fitting.

We cross-matched the CMaN members to the SPHEREx observations at their Gaia coordinates and extracted the available spectrophotometric measurements for each source using the `SPHEREx Photometry and Image Fitting Framework' (SPIFF) data reduction pipeline \citep{Gagne2026}\footnote{available at \url{https://github.com/jgagneastro/SPIFF}.}. SPIFF queries the IRSA SIA2 service using the proper motion-projected coordinates of each target at epoch 2025.5 to obtain a list of potentially eligible SPHEREx Level 2 images, then projects the coordinates of the target at each frame-specific epoch (using the FITS header key \texttt{MJD-AVG} of each eligible frame) to assemble a list of frames to be downloaded. SPIFF then retrieves a small detector cutout ($20\times\ 20$ pixels) from each frame, based on the expected target position, performs background subtraction based on the median value of the central $15\times\ 15$ pixels of the frame \footnote{after masking pixel flagged with bits 0, 1, 2, 6, 9, 10, 11, 15, 17, 19 and 21 as recommended by IRSA.}, and performs PSF fitting with \texttt{scipy.optimize}, using three free parameters (flux per steradian with a horizontal and vertical shift) to identify a preliminary optimal solution. The \texttt{UltraNest} \citep{Buchner2021} \footnote{available at \url{https://github.com/JohannesBuchner/UltraNest/}.} nested sampling algorithm is then used to explore the likelihood and determine posterior uncertainties. The saved tables used here pair the preliminary SciPy PSF flux estimates with these UltraNest uncertainties\footnote{Pixels flagged with bits 2, 6, 9, 10, 11, 15, and 17 are ignored during fitting, as recommended by IRSA.}. The flux per steradian is then converted to a flux measurement using the IRSA-provided median pixel solid angle (using the \texttt{HIERARCH OMEGA\_MEDIAN} key in the Level 2 FITS header) and the discrete sum of the detector-sampled PSF.

Multiple SPHEREx measurements of a given target were combined into a single spectral energy distribution after applying the channel-level quality filtering described in Section~\ref{sec:cuts}. The SPHEREx spectra were supplemented with Gaia DR3 $G_{\rm BP}$, $G$, and $G_{\rm RP}$ photometry and parallaxes \citep{Riello2021}, and with 2MASS $J$, $H$, $K_s$ \citep{Skrutskie2006} and AllWISE $W1$, $W2$ \citep{Wright2010} photometry from the Gaia DR3 best-neighbour cross-matches. Sources lacking the required Gaia or SPHEREx measurements were removed before fitting. After the availability requirements and the astrometric, contamination, and photometric quality cuts described below, the final SED sample contains 62 sources.

\section{Quality Cuts}
\label{sec:cuts}

Before fitting the combined Gaia and SPHEREx SEDs, we applied a series of astrometric, photometric, and source-confusion cuts designed to retain stars with reliable distances and minimally contaminated SPHEREx spectra. We queried Gaia DR3 for the source coordinates, $G$, $G_{\rm BP}$, and $G_{\rm RP}$ photometry, and the renormalized unit weight error (RUWE). To restrict the analysis to sources with sufficiently precise Gaia photometry, we required
\[
G<19.72~{\rm mag}.
\]

Because the SPHEREx angular sampling is substantially coarser than that of Gaia, unresolved neighboring sources can contaminate the extracted SPHEREx spectrum. We therefore searched Gaia DR3 for neighboring sources within $6\farcs15$ of each target, corresponding to the approximate SPHEREx pixel scale. For every neighbor with a measured Gaia $G$ magnitude, we estimated its flux relative to the target as
\begin{equation}
r_i=10^{-0.4(G_i-G_\star)},
\end{equation}
where $G_i$ and $G_\star$ are the neighbor and target magnitudes, respectively. The estimated fractional contamination from all identified neighbors was then calculated as
\begin{equation}
f_{\rm cont,G}=\frac{\sum_i r_i}{1+\sum_i r_i}.
\end{equation}
We excluded targets for which $f_{\rm cont,G}>0.05$. This contamination estimate is based on Gaia $G$-band flux ratios and is therefore an approximate diagnostic; neighbors with substantially different colors may contribute a different fractional contamination across the SPHEREx wavelength range.

To remove sources with potentially unreliable Gaia astrometry or unresolved multiplicity, we required ${\rm RUWE}\leq1.3$ and excluded sources without a finite RUWE measurement \citep{Penoyre26_position_resolvability}.

Finally, we excluded isolated outlying SPHEREx measurements. For each SPHEREx spectrum, we estimated the local expected flux from the mean of the adjacent channels,
\begin{equation}
F_{\lambda,{\rm local},j}=\frac{F_{\lambda,j-1}+F_{\lambda,j+1}}{2},
\end{equation}
such that a channel was excluded when
\begin{equation}
\frac{F_{\lambda,j}}{F_{\lambda,{\rm local},j}}>2
\qquad {\rm or} \qquad
\frac{F_{\lambda,j}}{F_{\lambda,{\rm local},j}}<\frac{1}{2}.
\end{equation}
This procedure removes isolated channels that deviate by more than a factor of two from their immediate neighborhood.

The SED sample contains 62 sources spanning approximately 157--211 pc and $G=7.6$--19.1. Three are hotter than the 7000 K limit of the atmosphere grid used for luminosities (Section~\ref{sec:sed_fitting}) and are not used further. A further seven fail the SPHEREx--photometry consistency test described in Section~\ref{sec:sed_fitting}. We then restricted the sample to PMS stars by requiring that sources lie more than 0.10 dex above the Feiden magnetic ZAMS and above the PARSEC ZAMS. The resulting HRD sample contains 29 stars at 157--204 pc, with $G=12.3$--17.5 and $T_{\rm eff}=3100$--4590 K. All 29 have published $P_{\rm sp}\geq0.872$ and $P_{\rm fin}\geq0.888$ (27 have $P_{\rm sp}>0.9$), and none has a negative radial-velocity or lithium membership flag. The published photometric membership selection and the ZAMS cuts can affect both the fitted locus and its width. Restricting the sample to $P_{\rm sp}>0.9$ changes the population-wide ages by at most 0.7 Myr (1.0 Myr for PARSEC; Section~\ref{sec:hrd_models}), and changing the ZAMS buffer to 0.05 or 0.15 dex leaves the sample unchanged.

The $G$-band neighbor cut is not a binary census or an infrared contamination limit. In a wider search, six HRD stars have an individual neighbor-to-target flux ratio above 0.05 in Gaia $G_{\rm RP}$ or 2MASS $K_s$ within $12\farcs3$. Excluding these sources gives a conservative 23-star control. Missing infrared counterparts do not imply zero contaminating flux, and summed flux at larger separations is not equivalent to flux entering the target PSF; the consistency test of Section~\ref{sec:sed_fitting} provides a direct, per-star check on contamination of the SPHEREx spectrum.

\section{SED Modeling and Parameter Inference}
\label{sec:models}
\label{sec:sed_fitting}

\begin{figure*}
\centering
\includegraphics[width=1\linewidth]{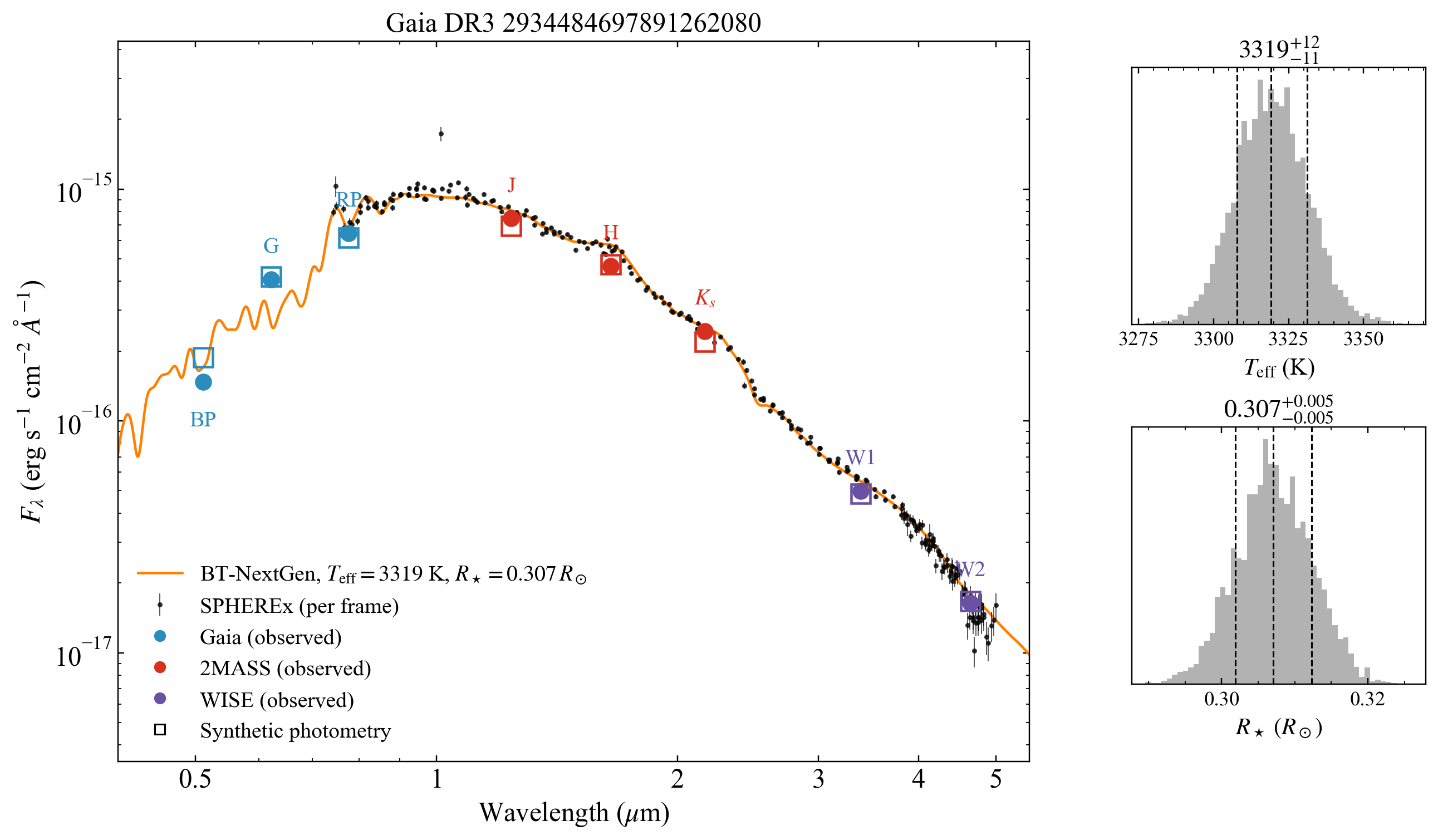}
\caption{Example SPHEREx spectral energy distribution and atmosphere-model fit for Gaia DR3 \texttt{\detokenize{2934484697891262080}}. Black points show the individual SPHEREx measurements with their reported uncertainties. Filled circles show the observed Gaia, 2MASS, and WISE photometry and open squares the synthetic photometry of the best-fitting BT-NextGen model. The orange curve shows that model at SPHEREx-like resolution, multiplied by the fitted gray SPHEREx factor $s$ (Section~\ref{sec:sed_fitting}). The right panels show the marginalized posteriors of $T_{\rm eff}$ and $R_\star$, with dashed lines marking the 16th, 50th, and 84th percentiles.}
\label{fig:sed_fit_example}
\end{figure*}

We modeled each quality-filtered source using the combined SPHEREx spectrum together with the Gaia, 2MASS, and WISE photometry. The stellar photosphere was represented with BT-NextGen model atmospheres \citep{allard2014} at solar metallicity, $[\alpha/{\rm Fe}]=0$, and fixed surface gravity $\log g=4.0$, spanning 2000--7000 K in 100 K steps. The model was scaled to the observed flux using the stellar radius and distance,
\begin{equation}
F_{\lambda,{\rm obs}}=F_{\lambda,{\rm surf}}\left(\frac{R_\star}{d}\right)^2.
\end{equation}
We fixed the extinction to $A_V=0$ mag due to the nearby and unreddened nature of the sample \citep{Kerr2025a}.

The atmosphere models were smoothed to $R=3000$ and linearly interpolated in temperature. Each SPHEREx measurement was modeled as the model spectrum averaged over a Gaussian bandpass centered on the measured wavelength, with a full width equal to the channel width reported by SPIFF for that frame ($R\approx35$--120). Averaging over the bandpass, rather than sampling the model at the channel center, changes the fitted temperatures by a median of 7 K. For the broadband measurements, we computed synthetic fluxes by integrating the model spectrum through the full response curve of each band,
\begin{equation}
\left\langle F_\lambda\right\rangle_{\rm band}
=
\frac{\int F_\lambda(\lambda)T(\lambda)\lambda\,d\lambda}
{\int T(\lambda)\lambda\,d\lambda},
\end{equation}
where $T(\lambda)$ is the band transmission function. Observed magnitudes were converted to fluxes using the Gaia DR3 zero points \citep{Riello2021}, the 2MASS zero points of \citet{Cohen2003}, and the WISE zero points of \citet{Jarrett2011}. We used the catalog flux uncertainties, with a 2\% floor for 2MASS and WISE.

For fitting the SPHEREx+Gaia data, we adopted a Gaussian likelihood of the form
\begin{equation}
\ln\mathcal{L}
=
-\frac{1}{2}
\sum_i
\left[
\frac{\left(F_{\lambda,i}^{\rm obs}-F_{\lambda,i}^{\rm mod}\right)^2}
{\sigma_{{\rm eff},i}^2}
+
\ln\left(2\pi\sigma_{{\rm eff},i}^2\right)
\right],
\end{equation}
where the effective uncertainty included both the reported measurement uncertainty and an additional fractional model uncertainty,
\begin{equation}
\sigma_{{\rm eff},i}^2
=
\sigma_i^2
+
\left(k_{\rm unc}F_{\lambda,i}^{\rm mod}\right)^2.
\end{equation}
The term $k_{\rm unc}$ accounts for imperfect atmosphere models, residual calibration differences, and variability between the SPHEREx and broadband epochs. Rather than fixing it, we fit $k_{\rm unc}$ for each star with a prior uniform in $\ln k_{\rm unc}$ between 0.001 and 0.5. For the HRD sample the fitted values have a median of 0.056 (16th--84th percentile range 0.048--0.067), and the reduced $\chi^2$ of the SPHEREx data is $\approx1.1$. Fixing $k_{\rm unc}=0.15$ instead changes the temperatures by a median of 4 K.

The SPHEREx and broadband fluxes are fit simultaneously, with the absolute flux scale set by the broadband photometry. We compared SPHEREx with 2MASS and WISE directly, by integrating the per-frame SPHEREx measurements through the $J$, $H$, $K_s$, and $W1$ passbands without any model. For the HRD sample the SPHEREx fluxes are 3.3\%, 4.3\%, 4.9\%, and 6.0\% brighter in these bands, with a star-to-star scatter of 0.015 dex (3.5\%) in the band-averaged ratio, which includes variability between epochs. To prevent this small offset from propagating into the luminosities, we multiply the SPHEREx model fluxes by a gray factor $s$, fit for each star with a prior uniform in $\ln s$ over $\pm0.7$, so that
\begin{equation}
F^{\rm mod}_{\lambda,i}=s\,F_{\lambda,{\rm surf}}(\lambda_i)\left(\frac{R_\star}{d}\right)^2
\end{equation}
for SPHEREx measurements and $s=1$ for the broadband photometry. The SPHEREx spectrum therefore constrains the temperature through its shape, while the absolute scale and luminosity are set by the Gaia, 2MASS, and WISE fluxes. The fitted factors have a median of $s=1.08$ (1.03--1.12). This is slightly larger than the direct infrared comparison because the atmosphere models are too bright in $G_{\rm BP}$ by $\approx0.09$ dex for the coolest stars; omitting $G_{\rm BP}$ changes the luminosities by a median of only 0.011 dex.

The gray factor also provides a direct test for contaminated or poorly extracted SPHEREx spectra, whose flux level is inconsistent with the broadband photometry. We flag the seven sources whose $\ln s$ lies more than three robust standard deviations from the sample median ($s<0.94$ or $s>1.26$) and exclude them from the HRD analysis.

The posterior distribution was sampled with the ensemble sampler \texttt{emcee} \citep{Foreman-Mackey2013}, using 32 walkers, 1500 burn-in steps, and 4000 retained steps. Temperature and radius priors are uniform, with $0.01<R_\star/R_\odot<5$, and the parallax has a Gaussian prior from Gaia. The retained chains exceed 80 autocorrelation times for every star, and 160 for every star in the HRD sample. We report the 16th, 50th, and 84th percentiles of the marginalized posterior distributions.

The stellar luminosity was calculated for every posterior sample using the bolometric surface flux of the atmosphere model integrated between 0.01~\micron\ and 75~\micron,
\begin{equation}
L_\star
=
4\pi R_\star^2 F_{\rm bol,surf}.
\end{equation}
Because the fitted radius is covariant with the sampled distance, the luminosity posterior incorporates the parallax uncertainty as a nuisance parameter as well as the uncertainties in temperature and flux normalization. For the HRD sample, the median formal uncertainties are 11 K in $T_{\rm eff}$ and 0.011 dex in $\log L/L_\odot$; the parallax contributes a median of 0.005 dex and at most 0.020 dex. These formal uncertainties do not include the absolute calibration of the broadband photometry or the atmosphere models. We treat a common $\pm0.02$ dex shift of the luminosity scale, comparable to the SPHEREx--2MASS--WISE offsets above, as a systematic uncertainty and propagate it to the ages in Section~\ref{sec:hrd_models}.

Figure~\ref{fig:sed_fit_example} shows an example fit for one CMaN member. We applied the same fitting procedure uniformly to all stars in the final quality-filtered sample and use the resulting marginalized temperature and luminosity uncertainties in the HRD analysis. The derived parameters for all 62 sources, including the fitted $s$ and $k_{\rm unc}$, the flags, and the membership probabilities, are listed in Table~\ref{tab:stellar}, which is available in machine-readable form.

\begin{figure}
\centering
\includegraphics[width=1\linewidth]{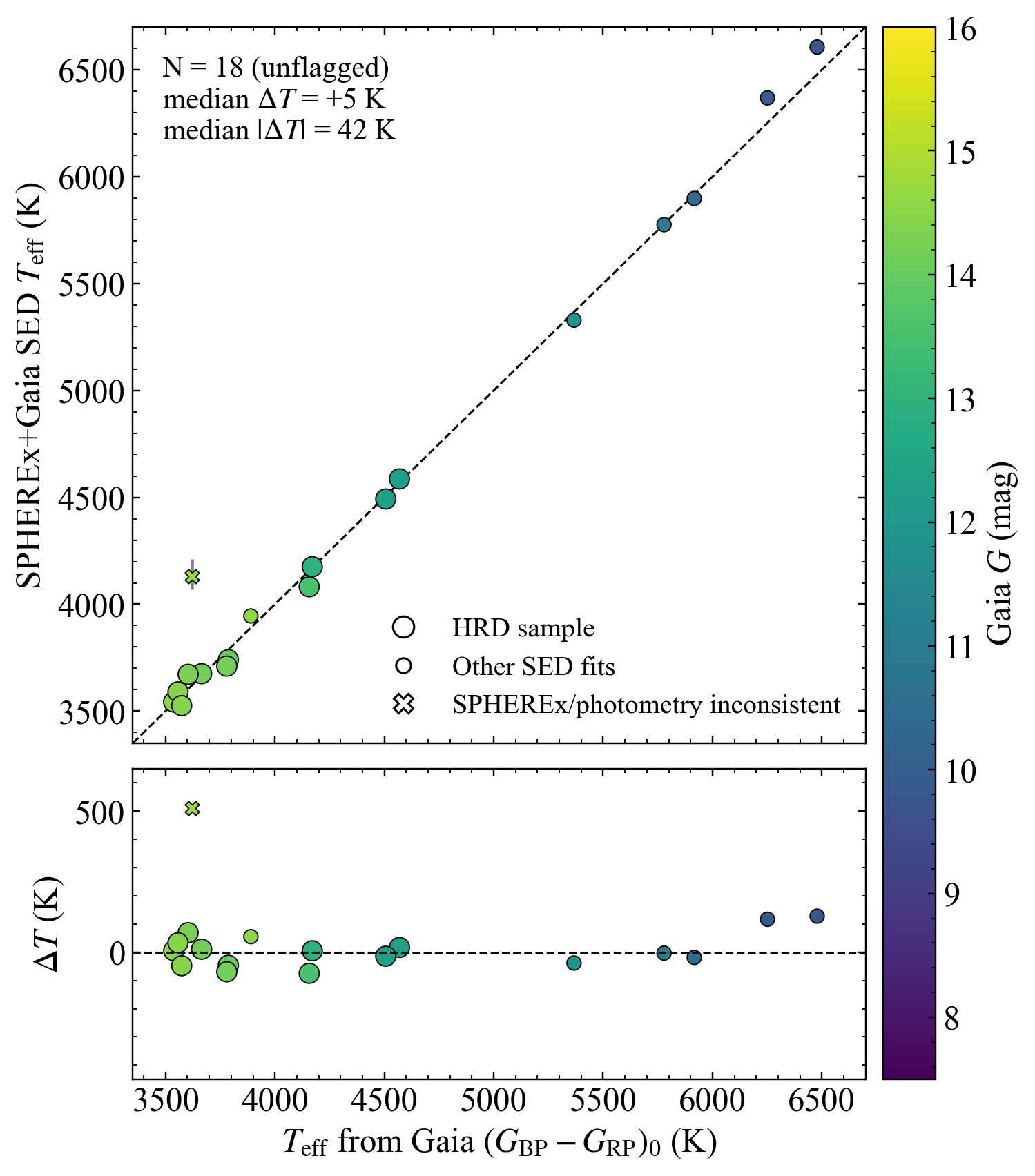}
\caption{SPHEREx+Gaia SED temperatures compared with those from the empirical Gaia $(G_{\rm BP}-G_{\rm RP})_0$ relation of \citet{Jeffries23}, for sources within the 3500--6500 K validity range of the color relation. Large symbols mark the HRD sample and the cross marks the source that fails the SPHEREx--photometry consistency test. The lower panel shows the difference (SED minus color temperature). Points are colored by Gaia $G$ magnitude.}
\label{fig:temperature_comparison}
\end{figure}

Fitting the SPHEREx data alone, without any Gaia, 2MASS, or WISE photometry, changes the temperatures of the HRD stars by a median of 14 K (maximum 33 K), and omitting $G_{\rm BP}$ changes them by 10 K (maximum 17 K). Adopting $\log g=4.5$ instead of 4.0 lowers them by a median of 25 K. Figure~\ref{fig:temperature_comparison} compares the SED temperatures with the empirical Gaia color relation of \citet{Jeffries23} used by \citet{Kerr2025a}, for the 19 stars within the 3500--6500 K validity range of that relation; fainter members are cooler than this range, which is why the figure contains only $G\lesssim16$ stars. For the 18 stars that pass the consistency test, the two scales agree to a median difference of $+5$ K and a median absolute difference of 42 K. The one flagged star in the figure, Gaia DR3 2947049710617661568 (3620 K from color versus 4130 K from the SED), has an SPHEREx flux 27\% above its broadband photometry, which indicates a contaminated extraction. A second discrepant source, Gaia DR3 2947050363452640640, has a color temperature of $\approx5700$ K but an SED fit outside the model grid, together with an anomalous extraction centroid, and is not used.

\begin{deluxetable*}{lcccccccc}
\tablecaption{Stellar parameters of the CMaN SED sample (excerpt)\label{tab:stellar}}
\tablehead{\colhead{Gaia DR3} & \colhead{$d$} & \colhead{$T_{\rm eff}$} & \colhead{$R_\star$} & \colhead{$\log L/L_\odot$} & \colhead{$s$} & \colhead{$k_{\rm unc}$} & \colhead{$P_{\rm sp}$} & \colhead{HRD} \\ \colhead{} & \colhead{(pc)} & \colhead{(K)} & \colhead{($R_\odot$)} & \colhead{} & \colhead{} & \colhead{} & \colhead{} & \colhead{}}
\startdata
2943161627037468800 & 170.6 & $3101^{+11}_{-11}$ & $0.331^{+0.006}_{-0.006}$ & $-2.045^{+0.015}_{-0.015}$ & 1.10 & 0.053 & 0.888 & Y \\
2933824956550574976 & 173.5 & $3208^{+13}_{-12}$ & $0.386^{+0.007}_{-0.007}$ & $-1.851^{+0.014}_{-0.013}$ & 1.08 & 0.067 & 0.991 & Y \\
2947002053651630080 & 179.5 & $3265^{+11}_{-11}$ & $0.645^{+0.009}_{-0.009}$ & $-1.374^{+0.011}_{-0.010}$ & 1.03 & 0.059 & 1.000 & Y \\
2948797624861727360 & 161.3 & $3449^{+10}_{-9}$ & $0.651^{+0.008}_{-0.007}$ & $-1.271^{+0.008}_{-0.008}$ & 1.09 & 0.048 & 0.999 & Y \\
2932320824649286656 & 169.6 & $3675^{+10}_{-9}$ & $0.763^{+0.008}_{-0.008}$ & $-1.022^{+0.008}_{-0.008}$ & 1.04 & 0.047 & 0.989 & Y \\
2946342037438710912 & 168.7 & $4586^{+12}_{-12}$ & $0.908^{+0.009}_{-0.009}$ & $-0.491^{+0.008}_{-0.008}$ & 1.08 & 0.048 & 0.984 & Y
\enddata
\tablecomments{Medians with 16th--84th percentile ranges. The full table of 62 sources, including Gaia, 2MASS, and AllWISE identifiers, consistency and grid flags, published membership probabilities, and lithium equivalent widths, is available in machine-readable form.}
\end{deluxetable*}

\section{Revised Lithium Age}
\label{sec:lithium_age}

We apply EAGLES \citep{Jeffries23} to the published lithium equivalent widths of \citet{Kerr2025a}, using our revised temperatures. The empirical relation includes the intrinsic equivalent-width scatter and yields a lithium-depletion age from stars across 3000--6500 K, rather than a lithium depletion boundary (LDB) measurement. The sample comprises the 14 stars that have published equivalent widths, SED fits within the 3000--6500 K calibration range, and no negative radial-velocity membership flag. Gaia DR3 2953470308761357312, which has zero equivalent width, a negative radial-velocity flag, and $P_{\rm fin}=0$, is excluded as a nonmember.

\begin{figure*}
\centering
\includegraphics[width=1\linewidth]{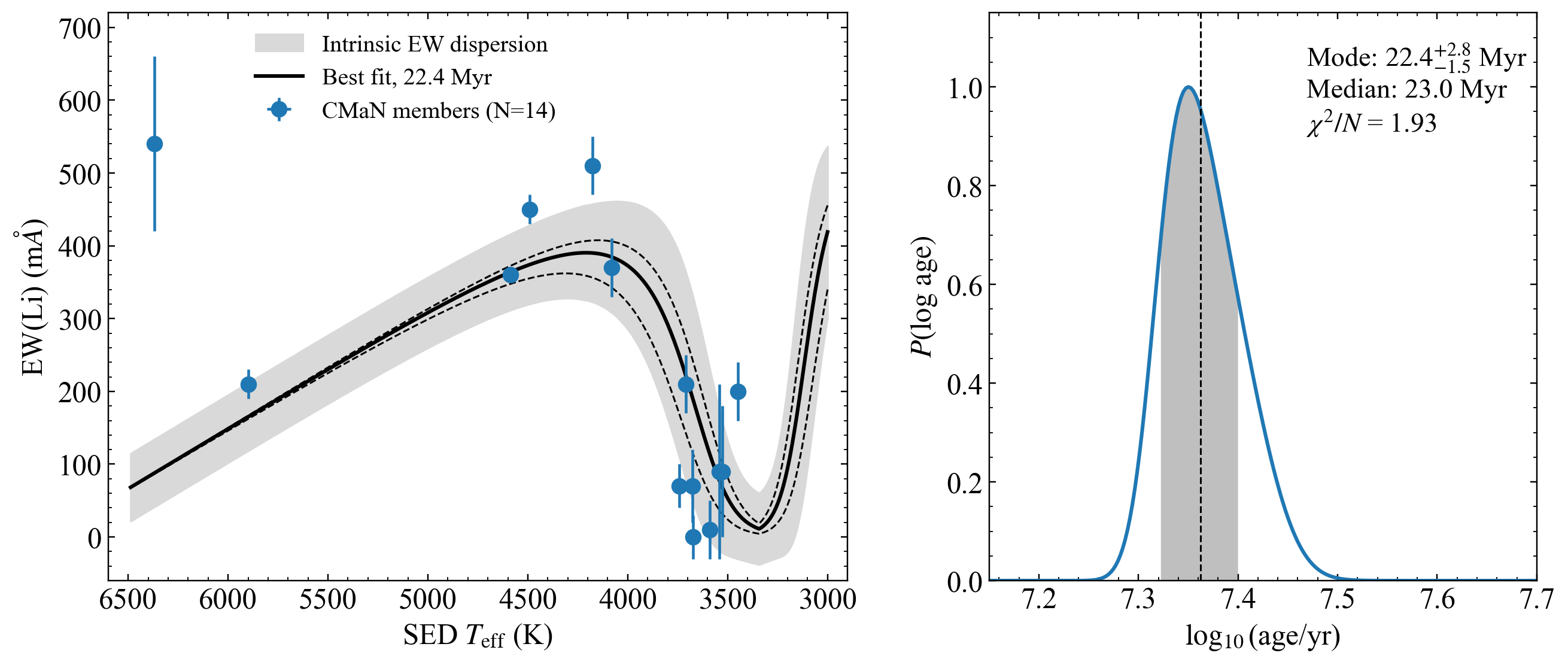}
\caption{EAGLES fit to the lithium equivalent widths of 14 membership-screened CMaN stars, using the SED temperatures. Left: observed equivalent widths and the best-fitting depletion sequence (solid), with the sequences at the 16th and 84th percentile ages (dashed) and the intrinsic equivalent-width dispersion at the preferred age (shaded). Right: posterior distribution of the population age; the shaded region marks the EAGLES 68\% interval and the dashed line the posterior median.}
\label{fig:spherex_lithium_eagles}
\end{figure*}

With a prior uniform in $\log_{10}(t/{\rm yr})$ between 6 and 8, the combined likelihood gives a modal age of $22.4^{+2.8}_{-1.5}$ Myr (Figure~\ref{fig:spherex_lithium_eagles}); the posterior median is 23.0 Myr, with $\chi^2/N=1.93$ including the intrinsic scatter. Omitting stars one at a time gives modal ages of 21.5--25.5 Myr, the largest change arising from the 3460 K member Gaia DR3 2948797624861727360. The age is insensitive to the SED choices tested in Section~\ref{sec:sed_fitting} (22.3--22.5 Myr). The lithium sample is limited mainly by SPHEREx coverage: of the 20 \citet{Kerr2025a} members ($P_{\rm fin}>0.5$) with lithium measurements in the calibrated temperature range, 14 have SED fits. Using all 20 gives $24.5^{+2.5}_{-2.0}$ Myr with SPHEREx temperatures where available and Gaia-color temperatures otherwise, and $26.6^{+2.2}_{-2.0}$ Myr with Gaia-color temperatures for all stars, close to the published value. This result agrees with the kinematic age of $23.7\pm4.9$ Myr. Lithium depletion is independent of the inferred luminosities, but it shares the SED temperature scale with the HRD analysis. We adopt the posterior median, 23.0 Myr, as the fixed age for the starspot tests in Section~\ref{sec:fspot_forward_model}.

\section{Testing Isochrone Fits with CMaN}
\label{sec:hrd_models}

\subsection{Population-Wide Isochronal Ages}

We compared the inferred $T_{\rm eff}$ and luminosities of the HRD sample with five sets of PMS evolutionary models: the magnetic and nonmagnetic models of \citet[][hereafter F16]{Feiden2016}, the SPOTS models of \citet{Somers2020}, the BHAC15 models \citep{Baraffe2015}, and the PARSEC v2.0 models \citep{Nguyen2022}, all at solar metallicity. For each grid we constructed isochrones from the native evolutionary tracks on a grid of ages uniform in $\log t$.

For a trial age $\tau$, each star is compared with the isochrone through the minimum covariance-weighted distance from the star to the curve. In addition to the common population age, we fit an intrinsic luminosity-scatter parameter, $\sigma_{\rm int}$, added in quadrature to the measured luminosity uncertainty,
\begin{equation}
\sigma_{\log L,{\rm tot},i}^2
=
\sigma_{\log L,i}^2+\sigma_{\rm int}^2.
\end{equation}
The likelihood includes the measured uncertainties in both $T_{\rm eff}$ and $\log L/L_\odot$. We adopt priors uniform in log age over 1--100 Myr and in log scatter over $0.001$--$0.3$ dex, and evaluate the two-dimensional posterior on a grid of 501 ages and 121 scatter values for each evolutionary grid; ensemble MCMC sampling of the same posterior reproduces the grid medians to within 0.1 Myr.

\begin{figure*}
\centering
\includegraphics[width=0.92\linewidth]{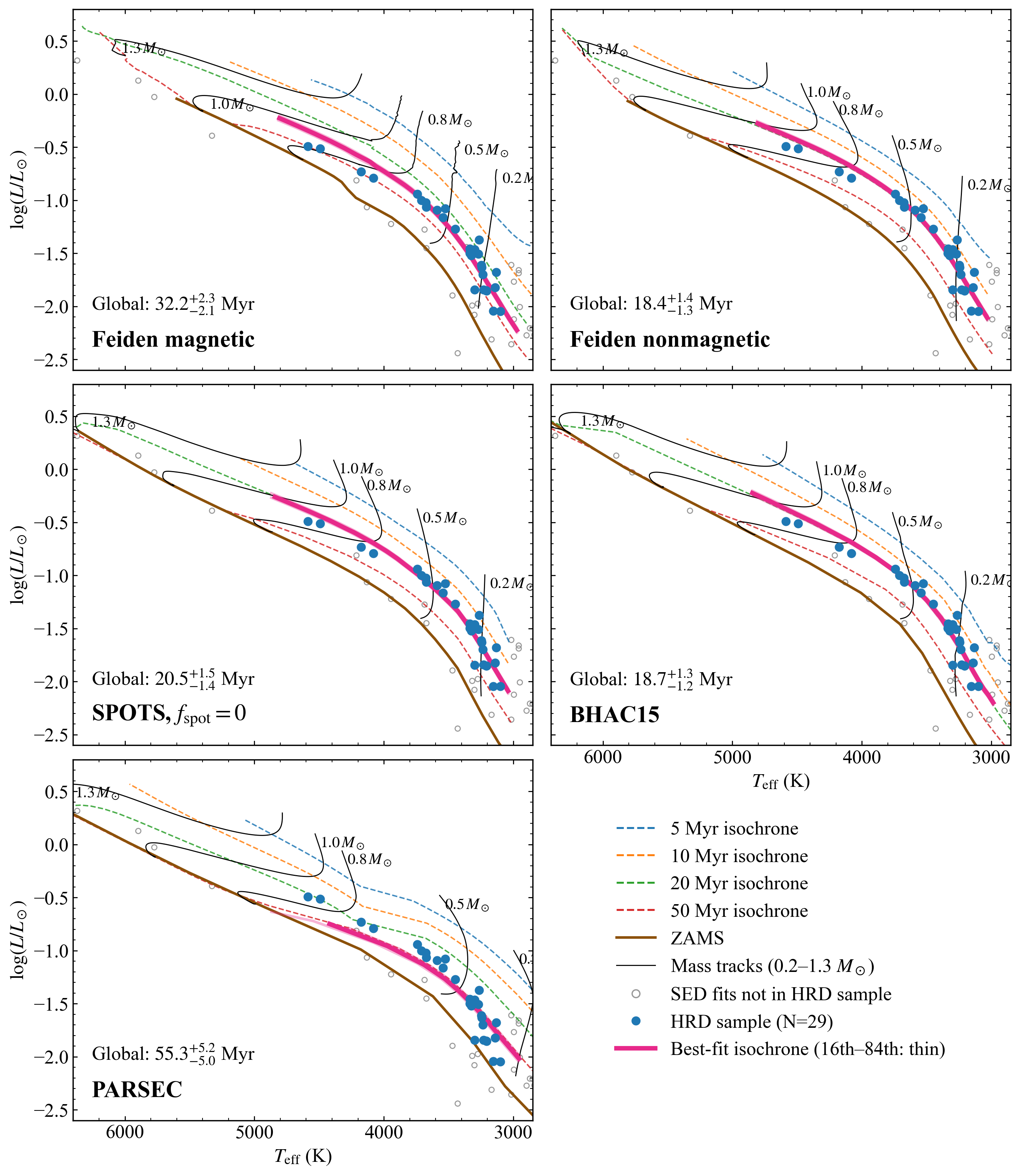}
\caption{HR diagrams for the CMaN HRD sample (blue) compared with five sets of PMS evolutionary models. Dashed curves mark the 5, 10, 20, and 50 Myr isochrones, the brown curve the ZAMS, and black curves representative mass tracks from $0.2$ to $1.3\,M_{\odot}$. Gray open circles show SED fits that are not in the HRD sample. The thick pink curve shows the best-fitting single isochrone for all stars, with the 16th and 84th percentile isochrones as thin curves; the corresponding age is given in each panel.}
\label{fig:hr_single}
\end{figure*}

Figure~\ref{fig:hr_single} shows the best-fitting isochrones, and Table~\ref{tab:global_fits} lists their ages and intrinsic luminosity dispersions. The unspotted SPOTS age, $20.5^{+1.5}_{-1.4}$ Myr, is consistent with the lithium and kinematic constraints. The Feiden nonmagnetic and BHAC15 ages, $18.4^{+1.4}_{-1.3}$ and $18.7^{+1.3}_{-1.2}$ Myr, are younger by $\approx4$ Myr, about $1.5\sigma$ from the lithium age, while Feiden magnetic and PARSEC require much older ages. The 18--55 Myr range among grids is much larger than the posterior uncertainty within any one grid.

\begin{table}
\centering
\caption{Population-wide isochronal fits (29 stars)}
\label{tab:global_fits}
\begin{tabular}{lcc}
\toprule
Grid & Age (Myr) & $\sigma_{\rm int}$ (dex) \\
\midrule
F16 magnetic & $32.2^{+2.3}_{-2.1}$ & $0.12^{+0.02}_{-0.02}$ \\
F16 nonmagnetic & $18.4^{+1.4}_{-1.3}$ & $0.11^{+0.02}_{-0.01}$ \\
SPOTS ($f_{\rm spot}=0$) & $20.5^{+1.5}_{-1.4}$ & $0.11^{+0.02}_{-0.01}$ \\
BHAC15 & $18.7^{+1.3}_{-1.2}$ & $0.11^{+0.02}_{-0.01}$ \\
PARSEC & $55.3^{+5.2}_{-5.0}$ & $0.16^{+0.02}_{-0.02}$ \\
\bottomrule
\end{tabular}
\end{table}

The Feiden nonmagnetic, BHAC15, and unspotted SPOTS isochrones pass near the center of the densely populated $3000$--$3800$ K sequence but lie above the warmer members, particularly at $4000$--$4700$ K. Their global ages therefore mainly reflect the cooler stars. Feiden magnetic shows a similar mismatch at an older overall age. PARSEC has the opposite curvature: its globally preferred isochrone passes through the sequence near $3400$--$3800$ K but lies above many of the coolest stars and below the warmer members. We fit separate temperature intervals to quantify these residuals.

The main observational systematics are the absolute luminosity scale and the surface gravity of the atmosphere models. A common $\pm0.02$ dex change of the luminosity scale moves every age by $\mp6$--7\%, and adopting the SPHEREx flux scale without the broadband anchor lowers the ages by $\approx10\%$, as expected from the 0.02 dex SPHEREx offset. Adopting $\log g=4.5$ in the SED fits lowers the ages by $\approx11\%$, whereas fitting $\log g$ freely leaves them unchanged. Membership, contamination screening, and the treatment of $G_{\rm BP}$ and $k_{\rm unc}$ change the F16 nonmagnetic, SPOTS, and BHAC15 ages by $\leq0.9$ Myr, the F16 magnetic age by $\leq1.4$ Myr, and the PARSEC age by $\leq2.2$ Myr. None of these choices changes the ordering of the grids.

\subsection{Temperature-Dependent Isochronal Ages}
\label{sec:temperature_dependent_ages}

\begin{figure*}
\centering
\includegraphics[width=0.92\linewidth]{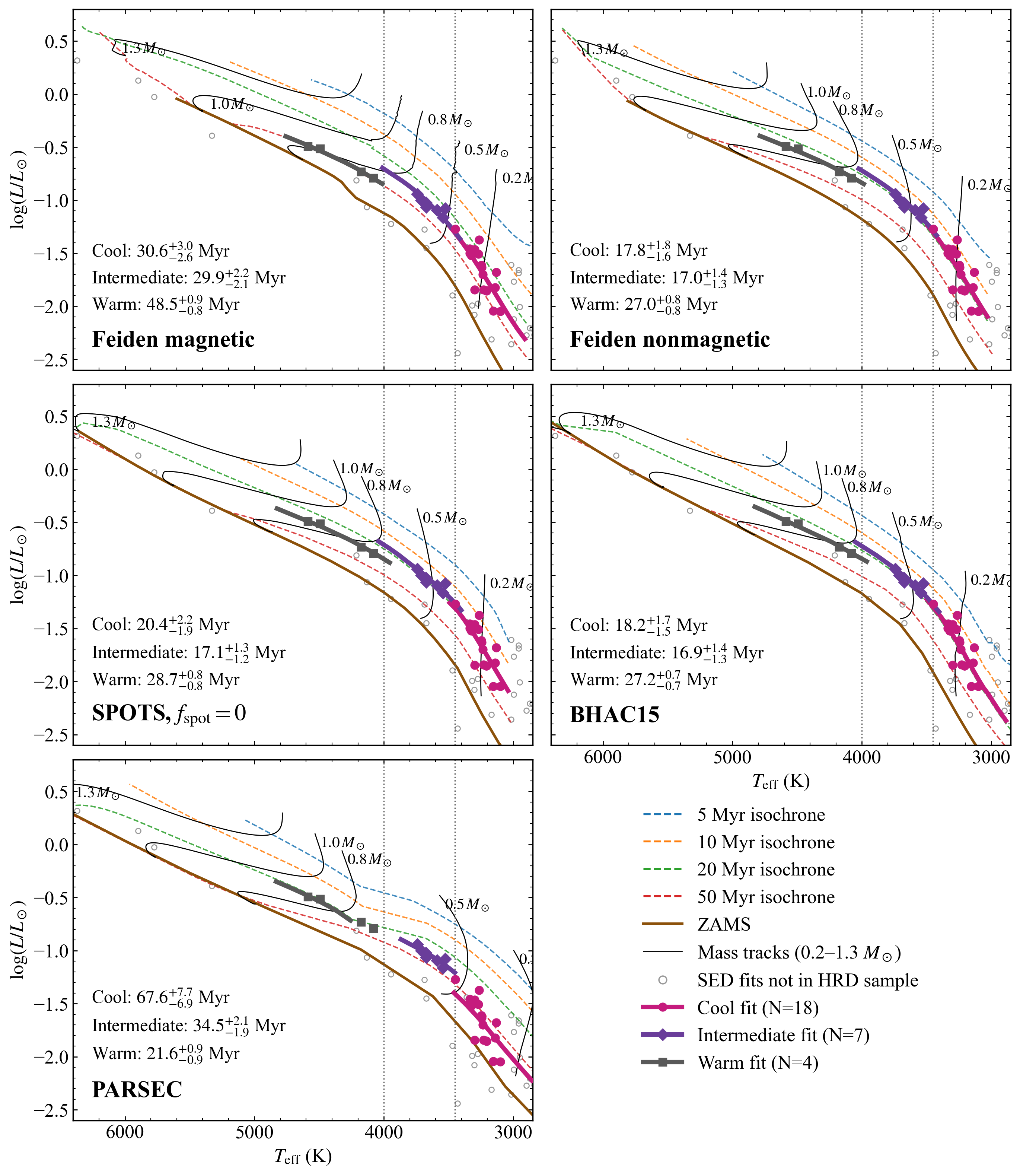}
\caption{As Figure~\ref{fig:hr_single}, but with the sample divided into cool, intermediate, and warm intervals at the temperatures marked by dotted lines, each fit independently with a single isochrone. The thick curves show the best-fitting isochrone of each interval over its temperature range, and the corresponding ages are given in each panel.}
\label{fig:temperature_hrd_fits}
\end{figure*}

We fit three temperature intervals independently: cool ($2500\leq T_{\rm eff}<3450$ K, 18 stars), intermediate ($3450\leq T_{\rm eff}<4000$ K, 7 stars), and warm ($4000\leq T_{\rm eff}\leq6300$ K, 4 stars). Table~\ref{tab:temperature_fits} lists the resulting ages, and Figure~\ref{fig:temperature_hrd_fits} shows the fits.

\begin{table*}
\centering
\caption{Temperature-dependent isochronal fits}
\label{tab:temperature_fits}
\begin{tabular}{lccccc}
\toprule
Grid & Cool age (Myr) & Intermediate age (Myr) & Warm age (Myr) & Cool $\sigma_{\rm int}$ (dex) & Interm. $\sigma_{\rm int}$ (dex) \\
\midrule
F16 magnetic & $30.6^{+3.0}_{-2.6}$ & $29.9^{+2.2}_{-2.1}$ & $48.5^{+0.9}_{-0.8}$ & $0.14^{+0.03}_{-0.02}$ & $0.06^{+0.02}_{-0.02}$ \\
F16 nonmagnetic & $17.8^{+1.8}_{-1.6}$ & $17.0^{+1.4}_{-1.3}$ & $27.0^{+0.8}_{-0.8}$ & $0.14^{+0.03}_{-0.02}$ & $0.05^{+0.02}_{-0.01}$ \\
SPOTS ($f_{\rm spot}=0$) & $20.4^{+2.2}_{-1.9}$ & $17.1^{+1.3}_{-1.2}$ & $28.7^{+0.8}_{-0.8}$ & $0.14^{+0.03}_{-0.02}$ & $0.04^{+0.02}_{-0.01}$ \\
BHAC15 & $18.2^{+1.7}_{-1.5}$ & $16.9^{+1.4}_{-1.3}$ & $27.2^{+0.7}_{-0.7}$ & $0.14^{+0.03}_{-0.02}$ & $0.05^{+0.02}_{-0.01}$ \\
PARSEC & $67.6^{+7.7}_{-6.9}$ & $34.5^{+2.1}_{-1.9}$ & $21.6^{+0.9}_{-0.9}$ & $0.15^{+0.03}_{-0.02}$ & $0.04^{+0.02}_{-0.01}$ \\
\bottomrule
\end{tabular}
\end{table*}

Feiden magnetic gives consistent cool and intermediate ages of 30--31 Myr, both older than the external constraints, and a warm age older by another $\approx18$ Myr. Feiden nonmagnetic, BHAC15, and unspotted SPOTS give 17--20 Myr below 4000 K, within 1--3$\sigma$ of the lithium age, but 27--29 Myr for the warm stars. PARSEC reverses the trend: its age increases from 22 Myr in the warm interval to 35 and 68 Myr in the intermediate and cool intervals. No grid returns a common age across the three intervals.

The intrinsic luminosity dispersion is concentrated among the coolest stars, where it reaches approximately $0.14$--$0.15$ dex, compared with $0.04$--$0.06$ dex at intermediate temperatures. The warm interval contains too few stars to constrain its intrinsic width.

\section{Can Starspots Explain the Observed Age Distributions?}
\label{sec:fspot_forward_model}

Starspots provide a natural mechanism for producing HRD dispersion within a coeval population because stars with different spot covering fractions occupy different temperature--luminosity loci at fixed mass and age \citep{Somers2015}. Large covering fractions are also observed empirically in individual weak-line T Tauri stars (WTTS). For example, the two-temperature analysis of LkCa~4 by \citet{Gully2017} found that a cool component covers approximately 80\% of the visible surface and demonstrated that accounting for this component substantially changes the star's inferred HRD position. Follow-up studies across broader populations of young stars have supported the apparent ubiquity of large spots on young stars \citep[e.g.,][]{Cao2022, Gangi, PerezPaolino2024, PerezPaolino2025separating}.

\subsection{Filling-Factor Estimates}
\label{sec:fspot_location}

We first determined which SPOTS filling factor brings each temperature interval into closest agreement with the adopted lithium and kinematic age of $23.0$ Myr. For each interval, we evaluated the likelihood across the available range in $f_{\rm spot}$ while holding the population age fixed and fitting an intrinsic luminosity scatter.

At 23.0 Myr the central loci correspond to $f_{\rm spot}=0.05^{+0.03}_{-0.03}$ for the cool interval and $0.16^{+0.04}_{-0.04}$ for the intermediate interval. The cool sequence still requires $\approx0.13$ dex of intrinsic luminosity dispersion. The warm posterior is weakly constrained ($0.05^{+0.12}_{-0.04}$), so we restrict the distribution experiments to the better-populated cooler intervals. Changing the fixed age to 18.8 or 28.6 Myr shifts the cool value from 0.02 to 0.12 and the intermediate value from 0.06 to 0.28.

These are values of the structural SPOTS parameter that align the central sequences with the lithium age. They need not equal spectroscopic covering fractions, which depend on the adopted spot contrast and geometry \citep[e.g.,][]{Gully2017}, and a central locus near the unspotted isochrone is compatible with substantial star-to-star variation around it. The spectroscopic measurements describe the visible spot coverage at the time of observation and allow both $f_{\rm spot}$ and $T_{\rm spot}/T_{\rm phot}$ to vary. The SPOTS evolutionary model calculations instead fix $T_{\rm spot}/T_{\rm phot}=0.8$ and hold the structural filling factor constant along each evolutionary track.

\subsection{Can a Different Age Accommodate Every Star?}
\label{sec:fspot_bracket}

\begin{figure*}
\centering
\includegraphics[width=0.85\linewidth]{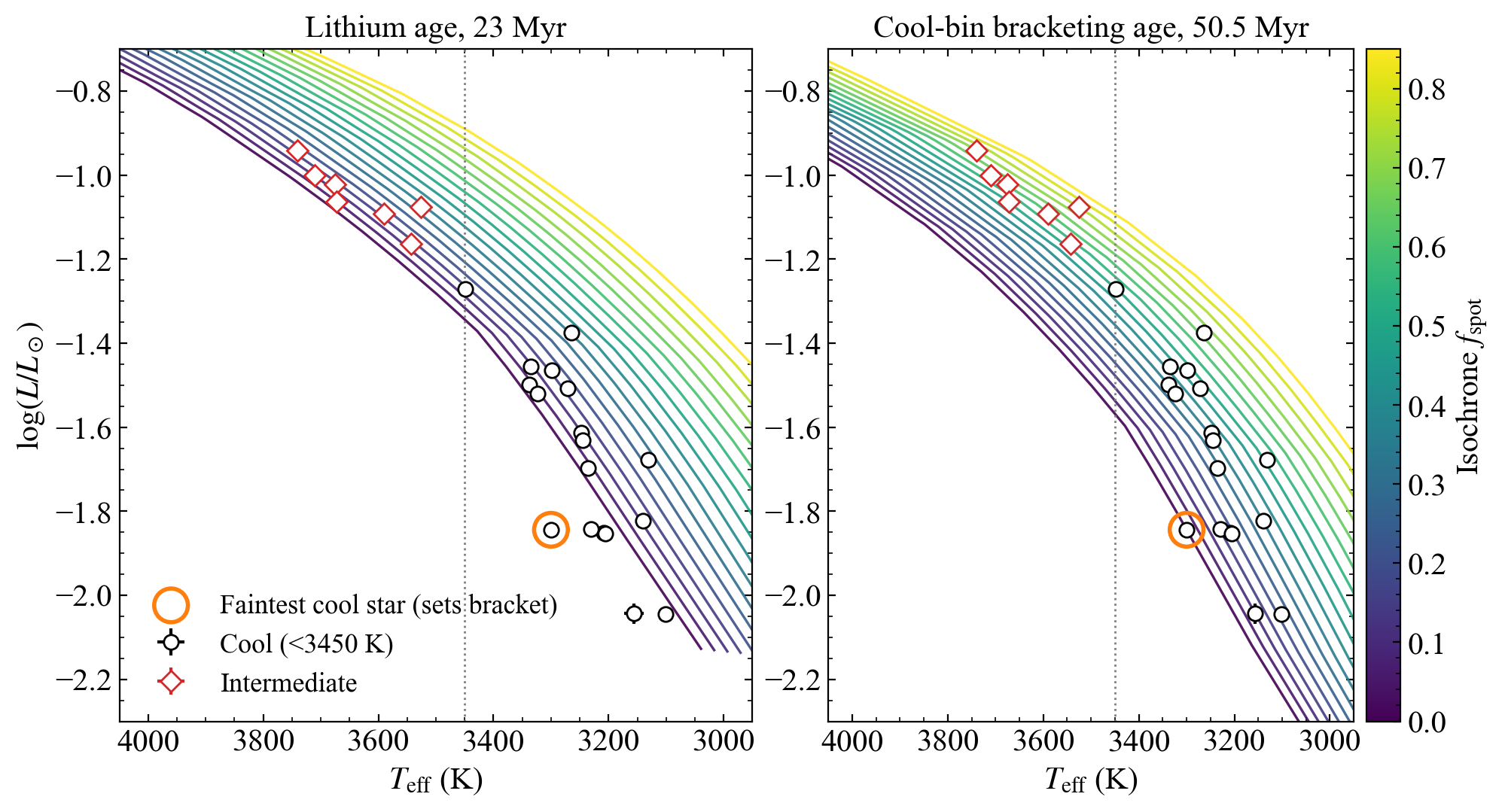}
\caption{Cool (circles) and intermediate (diamonds) CMaN stars compared with SPOTS isochrones for $0\leq f_{\rm spot}\leq0.85$ (colors) at the lithium age of 23.0 Myr (left) and at the 50.5 Myr age at which the unspotted isochrone passes through the faintest cool star (right; circled). The vertical dotted line marks the boundary between the two temperature intervals at $T_{\rm eff}=3450$ K.}
\label{fig:fspot_isochrone_overlay}
\end{figure*}

Figure~\ref{fig:fspot_isochrone_overlay} compares the cool and intermediate stars with SPOTS isochrones spanning $0\leq f_{\rm spot}\leq0.85$. Spots move a star's isochrone to higher luminosity at fixed temperature, so stars lying below the unspotted isochrone cannot be accommodated by any filling factor. At 23.0 Myr this applies to six of the 18 cool stars, whereas all seven intermediate stars fall within the spotted family, and all four warm stars lie below it.

The age could instead be chosen so that the unspotted isochrone passes through the faintest star of each interval, with the remaining stars assigned the filling factors that place them on spotted isochrones of that age. For the cool interval this bracketing age is 50.5 Myr, at which the cool stars require $f_{\rm spot}=0.35\pm0.15$ (median and standard deviation). The faintest cool star lies only 0.03 dex above the PARSEC ZAMS boundary of our sample selection; without it the bracketing age is 37 Myr, with $f_{\rm spot}=0.25\pm0.13$. The intermediate interval brackets at 21 Myr ($f_{\rm spot}=0.10\pm0.09$) and the warm interval at 29.5 Myr. All three intervals can be enclosed simultaneously only at ages of $\approx51$--54 Myr.

A hierarchical fit that treats the age as free and draws each star's filling factor from a bounded logistic-normal distribution, with no additional intrinsic scatter, leads to the same conclusion. The cool interval requires $62^{+9}_{-9}$ Myr ($45^{+13}_{-9}$ Myr without the faintest star), and a common age for the cool and intermediate intervals is $52^{+3}_{-4}$ Myr ($39^{+7}_{-4}$ Myr). An age that lets spots enclose the full cool sequence is therefore 1.6--2.8 times the lithium age, and it would make the cool stars older than the intermediate ones. We therefore retain the externally anchored age, at which the faint cool stars require a luminosity-lowering effect not represented by starspots in SPOTS, such as residual measurement errors, unresolved companions affecting the SED, or a finite age spread.

\subsection{Forward Modeling Using Empirical Spot Distributions for Young Stars}
\label{sec:fspot_empirical}

We next tested whether star-to-star variations in spot coverage can account for the luminosity dispersion of the cool and intermediate CMaN sequences at a fixed age of $23.0$ Myr. \citet{PerezPaolino2025b} measured spot filling factors from high-resolution optical--near-infrared spectra for a sample of weak-line T Tauri stars. We use population summaries of those measurements to test whether the corresponding structural heterogeneity can reproduce the observed luminosity scatter around the 23.0 Myr isochrone.

We combine the empirical population standard deviations, 0.223 for 26 stars below 3500 K and 0.109 for 12 stars between 3500 and 4000 K, with adopted structural medians of 0.28 and 0.75 in bounded logistic-normal distributions.

We drew population samples using a Monte Carlo algorithm. In each iteration, we drew one filling factor for every star and evaluated
\begin{equation}
T_{{\rm eff},i}^{\rm mod}
=
T_{\rm eff}\left(M_i,\tau,f_{{\rm spot},i}\right)
\end{equation}
and
\begin{equation}
\log L_i^{\rm mod}
=
\log L\left(M_i,\tau,f_{{\rm spot},i}\right),
\end{equation}
with $\tau=23.0$ Myr. We then perturbed the model predictions using the individual temperature and luminosity uncertainties while retaining the observed 18- and seven-star memberships.

Luminosity residuals were measured relative to the corresponding reference relation,
\begin{equation}
\Delta\log L_i
=
\log L_i-\log L_{\rm ref}\left(T_{{\rm eff},i}\right).
\end{equation}
We characterized the residual distribution using its standard deviation and asymmetry,
\begin{equation}
A
=
\left(P_{84}-P_{50}\right)
-
\left(P_{50}-P_{16}\right).
\end{equation}
We generated 20,000 realizations for each temperature interval. Control simulations assigned every star the same characteristic filling factor and therefore included observational uncertainties and finite sample size, but no intrinsic variation in $f_{\rm spot}$.

The results are summarized in Table~\ref{tab:forward_widths}. For the cool population, the observed residual width is 0.142 dex. A common filling factor predicts 0.023 dex, and none of the 20,000 realizations reaches the observed width (one-sided 95\% upper bound on the exceedance probability $\approx1.5\times10^{-4}$). The empirical distribution predicts 0.201 dex and reaches or exceeds the observed width in 95\% of realizations, with a positive median asymmetry, $A=0.123$ dex, opposite to the observed $A=-0.079$ dex. For the intermediate population, the observed width of 0.037 dex is reproduced by the empirical distribution (median 0.034 dex; exceedance probability 40\%), whereas a common filling factor gives 0.013 dex and never reaches the observed width. The observed asymmetry, $A=0.003$ dex, is consistent with the variable model's median of $-0.008$ dex. The empirical distribution is therefore compatible with the intermediate sequence but overpredicts both the width and the positive tail of the cool sequence.

\begin{table}
\centering
\caption{Forward-model luminosity widths at 23.0 Myr}\label{tab:forward_widths}
\begin{tabular}{lcccc}
\toprule
Interval & $N$ & Observed & Common $f$ & Variable $f$\\
 & & (dex) & (dex) & (dex) \\
\midrule
Cool & 18 & 0.142 & 0.023 & 0.201 (95\%)\\
Intermediate & 7 & 0.037 & 0.013 & 0.034 (40\%)\\
\bottomrule
\end{tabular}
\tablecomments{Median simulated widths; parentheses give the fraction of variable-$f$ realizations at least as wide as observed. No common-$f$ realization reaches the observed width.}
\end{table}

Using Gaia-ESO observations, \citet{Franciosini22_gaia-eso_survey} found that models with approximately 20\% effective spot coverage improved the simultaneous agreement with the CMD and lithium depletion patterns of several young clusters. A single common covering fraction was nevertheless insufficient to explain the full lithium dispersion, with greater coverage toward lower masses appearing to be required. The greater structural heterogeneity required for the low-temperature CMaN population is qualitatively consistent with the broader conclusion that a single spot prescription need not describe all members of a young population.

\section{Discussion}\label{sec:discussion}
\subsection{Mass-dependent Performance of PMS Evolutionary Models}
No grid returns a common age across the CMaN temperature range (Table~\ref{tab:temperature_fits}). The pattern found with the Feiden nonmagnetic, BHAC15, and unspotted SPOTS grids, cool stars near the external age and warmer stars older, resembles the classical Upper Sco discrepancy \citep{Pecaut2016,Fang2017}, which has been attributed to some combination of convection prescriptions, atmospheric boundary conditions, temperature scales, magnetic activity, and unresolved multiplicity \citep[e.g.,][]{DaRio2010,Bell12_pre-main-sequence_isochrones,Bell14_pre-main-sequence_isochrones,SullivanKraus2021}. CMaN shows that the discrepancy persists to $\sim23$ Myr, when accretion and disk-related excesses have long since faded and cannot contribute to it.

Magnetic models were developed in part to address this problem. Inhibition of convection inflates the radius, lowers the effective temperature, and slows PMS contraction, so magnetic models assign older ages to low-mass stars. In Upper Sco, \citet{Feiden2016} found that magnetic models gave a consistent $\approx10$ Myr age from A through M spectral types. In CMaN, the Feiden magnetic grid is older than the external constraints in all three intervals and retains the age increase above $4000$ K. Two features of the F16 calculations may contribute. The surface field is set by photospheric equipartition, giving $\approx2.5$ kG, rather than by the rotation of each star; this is appropriate for saturated dynamos (${\rm Ro}\lesssim0.1$; \citealt{Reiners2009}), but warmer stars with thinner convective envelopes may leave saturation earlier \citep{See2018} and would then be assigned too strong a field. In addition, the structural response to a given field changes as the radiative core grows \citep{Feiden2016}. Distinguishing these effects requires rotation periods and field measurements for the individual stars.

Dynamical masses point to the same conclusion: magnetic and spot physics improve agreement in some mass ranges and degrade it in others \citep{Braun2021,Towner2025,Towner2026,Zallio2026}. The older Feiden magnetic ages in CMaN constrain this particular prescription of the surface field, not the importance of magnetic effects in general.

In the temperature--luminosity plane, BHAC15 gives ages within 0.5 Myr of Feiden nonmagnetic in every interval, whereas \citet{Kerr2025a} found a BHAC15 color--magnitude age of $13\pm0.3$ Myr. Because our comparison avoids synthetic colors and bolometric corrections, this suggests that the young BHAC15 color--magnitude ages arise mainly from the transformation of the models into the Gaia bands rather than from their interior structure.

PARSEC behaves differently because its low-mass models include an empirical boundary condition. \citet{Chen14_improving_parsec} found that models with PHOENIX BT-Settl $T$--$\tau$ relations predicted radii too small and optical sequences too faint and blue, and added an ad hoc shift to the boundary relation increasing from zero at $4730$ K to $\approx14\%$ at $3160$ K, calibrated on main-sequence dwarf radii \citep[see also][]{Nguyen2022}. The steep increase of the PARSEC age toward low temperatures in CMaN, from 22 Myr above 4000 K to 68 Myr below 3450 K, indicates that this dwarf calibration does not transfer to a $\sim23$ Myr PMS population. Because the discrepancy appears directly in the temperature--luminosity plane, it cannot arise solely from the transformation into Gaia colors.

\subsection{What SPHEREx Adds to Broadband Photometry}
\label{sec:spherex_value}
Because the absolute flux scale of our fits is set by the Gaia, 2MASS, and WISE photometry (Section~\ref{sec:sed_fitting}), the contribution of SPHEREx lies in the temperatures rather than the luminosities. To quantify it, we refit every source with the same atmosphere grid, priors, and fitted model term, but using only the eight broadband measurements. For the HRD sample, the luminosities are unchanged (median difference $+0.006$ dex), as expected when the bolometric flux is anchored by the same photometry. The temperatures are not. Without SPHEREx, the median formal temperature uncertainty increases from 11 to 27 K, and the temperatures shift systematically cooler, by a median of 90 K for the full sample and 99 K for the stars below 3450 K.

The shift is a bias of the photometry-only fits rather than of the SPHEREx fits. Compared with the empirical Gaia-color scale of \citet{Jeffries23}, the full SED temperatures agree to a median of $+5$ K (Figure~\ref{fig:temperature_comparison}), whereas the photometry-only temperatures are $49$ K cooler for the same 18 stars. Without SPHEREx, the temperature of an M star is constrained mainly by the optical bands, where the atmosphere models fit the coolest stars poorly (the model $G_{\rm BP}$ flux is $\approx0.09$ dex too bright; Section~\ref{sec:sed_fitting}). With SPHEREx, the temperature is instead set by the shape of the near-infrared spectrum, including its molecular bands: fitting the SPHEREx data alone reproduces the full-fit temperatures to a median of 14 K.

This difference propagates directly into the ages. At fixed luminosity, a 90 K shift along the nearly vertical PMS tracks of M stars moves the inferred mass and age substantially. Re-applying our sample selection to the photometry-only parameters (28 stars) gives population-wide ages of 13.2, 14.7, 13.5, 24.1, and 42.9 Myr for Feiden nonmagnetic, SPOTS, BHAC15, Feiden magnetic, and PARSEC, $\approx20$--30\% younger than the adopted values in Table~\ref{tab:global_fits} and well below the lithium age. Several of the coolest members, at $\approx3025$ K, would also fall below the low-mass limit of the SPOTS grid at most ages. The ordering of the grids and the opposite temperature trends of Section~\ref{sec:temperature_dependent_ages} persist without SPHEREx, but the absolute age scale does not.

SPHEREx therefore contributes most where broadband SEDs are weakest: the temperatures of M stars, which dominate the membership of nearby young associations and the leverage of population-wide fits. Because the survey covers the full sky with the same spectral sampling, the same calibration can be applied homogeneously to every nearby association, without the target-by-target spectroscopy usually required to place low-mass stars on the HRD.

\subsection{Implications for Isochronal Age Measurements}
The posterior uncertainties of individual fits, 1--8 Myr, are small compared with the 18--55 Myr range among grids. They measure precision conditional on a model; they exclude the systematic uncertainty of the evolutionary physics \citep{Soderblom2014,Herczeg15}. The observational systematics that we can quantify, dominated by the absolute luminosity scale and the gravity of the atmosphere models, amount to 6--11\%. For CMaN, the choice of grid is therefore the dominant source of uncertainty in the absolute age.

A population-wide age is also a weighted average over the sampled mass range. The likelihood is dominated by the most populated parts of the HRD and by the stars with the smallest uncertainties. In CMaN, the cool members pull the Feiden nonmagnetic, BHAC15, and SPOTS ages toward the external benchmark despite the older ages of the warm stars, and they pull PARSEC to an age much older than that of its warm members. Two coeval associations analyzed with the same grid can therefore receive different ages if their samples cover different temperature ranges. More distant populations have fewer detected low-mass members, so, if the CMaN trends hold, PARSEC would assign older ages to nearby, more complete samples, and the Feiden, BHAC15, and SPOTS grids would show the opposite bias \citep[see also][]{Wang25}. Mass-dependent binary fractions and detectability can introduce a similar gradient \citep{SullivanKraus2021}. Relative ages are less sensitive to a common model zero point \citep{Bell12_pre-main-sequence_isochrones,Bell13_pre-main-sequence_isochrones,Herczeg15}, but mass-dependent errors do not cancel between samples with different temperature coverage. Isochronal ages should therefore be reported with the grid and the fitted temperature range, and comparisons among associations should use a common interval \citep{Kerr2025a}.

Below $4000$ K, the unspotted SPOTS age of the cool stars agrees with the lithium age, while its intermediate-temperature age and the Feiden nonmagnetic and BHAC15 ages are 2--3$\sigma$ younger. For M stars at $\sim20$ Myr, Gaia and SPHEREx SEDs combined with these grids therefore recover ages within $\approx25\%$ of the lithium scale. Whether the agreement holds at other ages remains to be tested, because spot coverage and magnetic activity evolve during the PMS.

\subsection{Spot Heterogeneity and Apparent Age Spreads}
The location and width of the cool sequence carry different information. At the lithium age its central locus lies close to the unspotted SPOTS isochrone (Section~\ref{sec:fspot_location}), yet its width requires star-to-star differences in the structural filling factor if spots are the only cause, smaller than those implied by the empirical distribution of young stars (Section~\ref{sec:fspot_empirical}). Its faint edge cannot be produced by spots at all (Section~\ref{sec:fspot_bracket}).

These inferences are conditional on the SPOTS assumptions of a fixed spot-to-photosphere temperature ratio, $T_{\rm spot}/T_{\rm phot}=0.8$, and a filling factor that remains constant along each track \citep{Somers2020}. Neither is expected to hold in general. Spot temperature contrasts vary among stars and with photospheric temperature \citep{Berdyugina2005,PerezPaolino2024,PerezPaolino2025separating}, so the same covering fraction need not block the same fraction of the luminosity across the CMaN mass range. Spot coverage should also evolve with rotation, field strength, and internal structure \citep{Morris2020,Cao2023,Cao2025}. A fixed-$f_{\rm spot}$ track therefore describes a star subjected to persistent flux blocking rather than a self-consistent magnetic history. These differences also explain why spectroscopic filling factors, which describe the instantaneous visible surface with free spot temperatures, need not transfer into the structural SPOTS parameter.

\subsection{Limitations}
The main limitation is the sample size after the quality and ZAMS cuts, which leaves four stars in the warm interval and makes its age sensitive to individual stars. The SED fits adopt a single-temperature atmosphere at fixed $\log g=4.0$ and $A_V=0$. The fitted $T_{\rm eff}$ is therefore an effective temperature of a spotted, multi-temperature surface \citep{Gully2017,PerezPaolino2024}. Accretion history and initial entropy can alter both HRD positions and lithium depletion \citep{Baraffe17_self-consistent_evolution}; the grids used here do not vary them.

Our PMS-selected sample excludes stars near the ZAMS and provides no turnoff age. \citet{Kerr2025a} identify two bright Hipparcos stars as candidate members, including Mirzam ($\beta$~CMa), a $13.5\,M_\odot$ $\beta$~Cephei pulsator with an asteroseismic age of $12.4\pm0.7$ Myr \citep{Mazumdar2006}. If Mirzam is a coeval member, its age is about half the lithium and kinematic ages. Its membership rests on Hipparcos astrometry alone, and the age of a star of this mass depends on the adopted core overshooting and rotation, so we do not use it as a constraint. A radial velocity for Mirzam and an analysis of the second candidate, HIP~32848, would test whether CMaN contains an evolved massive population. Larger SPHEREx samples spanning a range of ages will allow the mass-dependent residuals and spot-related dispersions found here to be tested more broadly.

\section{Conclusions}
\label{sec:conclusions}

We used Gaia+SPHEREx SEDs, anchored to 2MASS and WISE photometry, together with lithium equivalent widths and published kinematics, to test five PMS evolutionary grids against CMaN. We find:
\begin{enumerate}
\item The SPHEREx fluxes agree with 2MASS and WISE to 3--6\%. The revised EAGLES lithium age is $22.4^{+2.8}_{-1.5}$ Myr, consistent with the dynamical age of $23.7\pm4.9$ Myr.
\item Population-wide isochronal ages span $18$--$55$ Myr. Unspotted SPOTS ($20.5^{+1.5}_{-1.4}$ Myr) agrees with the external age, Feiden nonmagnetic and BHAC15 (18--19 Myr) are slightly younger, and Feiden magnetic and PARSEC require older ages. Observational systematics shift these ages by 6--11\%.
\item Every grid produces a temperature-dependent age. Feiden, BHAC15, and SPOTS isochrones make stars above $4000$ K older than the cooler members, while PARSEC isochrones give the opposite trend.
\item The coolest members have excess luminosity dispersion that a common SPOTS filling factor cannot reproduce. The empirical filling-factor distribution of young stars overpredicts this width (0.20 versus 0.14 dex) but reproduces the intermediate-temperature sequence. Spotted isochrones can enclose every cool member only at 37--50 Myr, inconsistent with the lithium and kinematic ages.
\end{enumerate}

\begin{acknowledgments}
This publication makes use of data products from the Two Micron All Sky Survey, which is a joint project of the University of Massachusetts and the Infrared Processing and Analysis Center/California Institute of Technology, funded by NASA and the NSF, and from the Wide-field Infrared Survey Explorer, which is a joint project of the University of California, Los Angeles, and the Jet Propulsion Laboratory/California Institute of Technology, funded by NASA. Anthropic Claude was used to assist with preparation of analysis scripts. Responsibility for the scientific content and final manuscript rests with the authors.
\end{acknowledgments}

\bibliography{spots_working}
\bibliographystyle{aasjournalv7}

\end{document}